\documentclass[cits]{PoS}
\pdfoutput=1
\usepackage{multicol}
\usepackage{slashed}
\usepackage{setspace}
\usepackage{color}
\usepackage{booktabs}
\usepackage{epstopdf}
\usepackage[pdftex]{graphicx}
\usepackage[symbol*,hang,flushmargin]{footmisc}
\title{Matrix Elements for $D$- and $B$-Mixing from 2+1 Flavor Lattice QCD}

\ShortTitle{Matrix Elements for $D$- and $B$-Mixing from 2+1 Flavor Lattice QCD}

\author{\speaker{C.C. Chang} \textsuperscript{\textnormal{\textit{a f}}}, C. Bernard \textsuperscript{\textnormal{\textit{b}}}, C.M. Bouchard \textsuperscript{\textnormal{\textit{c}}}, A.X. El-Khadra \textsuperscript{\textnormal{\textit{a~f}}}, E.D. Freeland \textsuperscript{\textnormal{\textit{d}}}, E.~G\'amiz \textsuperscript{\textnormal{\textit{e}}}, A.S. Kronfeld \textsuperscript{\textnormal{\textit{f}}}, J. Laiho \textsuperscript{\textnormal{\textit{g}}}, R.S. van de Water \textsuperscript{\textnormal{\textit{f}}}\\
        \textsuperscript{a}Physics Department, University of Illinois, Urbana, IL 61801, USA\\
				\textsuperscript{b}Department of Physics, Washington University, St. Louis, MO 63130, USA\\
				\textsuperscript{c}Department of Physics, The Ohio State University, Columbus, OH 43210, USA\\
				\textsuperscript{d}Liberal Arts Department, The School of the Art Institute of Chicago, Chicago, IL 60603, USA\\
				\textsuperscript{e}CAFPE and Departamento de Fisica Teorica y del Cosmos, Universidad de Granada, E-18002 Granada, Spain\\
				\textsuperscript{f}Theoretical Physics Department, Fermi National Accelerator Laboratory\footnote{Fermilab is operated by Fermi Research Alliance, LLC, under Contract No. DE-AC02-07CH11359 with the United States Department of Energy}, Batavia, IL 60510, USA\\
				\textsuperscript{g}SUPA, Department of Physics and Astronomy, University of Glasgow, Glasgow, G12-8QQ, United Kingdom\\}
\author{Fermilab Lattice and MILC Collaborations\\
        E-mail: \email{cchang5@illinois.edu}}

\abstract{We present the status of our calculation of hadronic matrix elements for $D$- and $B$-meson mixing. We use a large set of the MILC collaboration's $N_f=2+1$ asqtad ensembles, which includes lattice spacings in the range $a\approx0.12$--0.045~fm, and up/down to strange quark mass ratios as low as 0.05. The asqtad action is also employed for the light valence quarks. For the heavy quarks we use the Sheikholeslami-Wohlert action with the Fermilab interpretation.  Our calculation covers the complete set of five local operators needed to describe $B$-meson mixing in the Standard Model and Beyond. In the charm sector, our calculation of local mixing matrix elements may be used to constrain new physics models. We present final correlator fit results on the full data set for the $B$-meson mixing project and preliminary fit results for the $D$-meson mixing project.}

\FullConference{31st International Symposium on Lattice Field Theory - LATTICE 2013\\
		July 29 - August 3, 2013\\
		Mainz, Germany}

\begin{document}
\section{Neutral meson mixing phenomenology}
The mass and width differences in a neutral meson system are described by the parameters $M_{12}$ and $\Gamma_{12}$, which in turn can be related to matrix elements of the effective weak Hamiltonian as follows:
\begin{equation}
M_{12}+\frac{i}{2}\Gamma_{12}=\left<\bar{H}\left|\mathcal{H}_W^{\Delta h=2}\right|H\right>+\sum_{f}\frac{\left<\bar{H}\left|\mathcal{H}_W^{\Delta h=1}\right|f\right>\left<f\left|\mathcal{H}_W^{\Delta h=1}\right|H\right>}{E_f-M_H}. \label{mixingH}
\end{equation}
where $H$ represents an $h$-flavored heavy-light meson with $h=b$, or $c$. $M_{12}$ receives contributions from both terms in Eq.~(\ref{mixingH}). The $\mathcal{H}_W^{\Delta h=2}$ term describes the short-distance mixing contributions while the $\mathcal{H}_W^{\Delta h=1}$ is the long-distance contribution mediated by on-shell intermediate states. $\Gamma_{12}$ receives contributions from the $\mathcal{H}_W^{\Delta h=1}$ term mediated by off-shell intermediate states.

\begin{figure}[h]
	\centering
		\includegraphics[width=\linewidth]{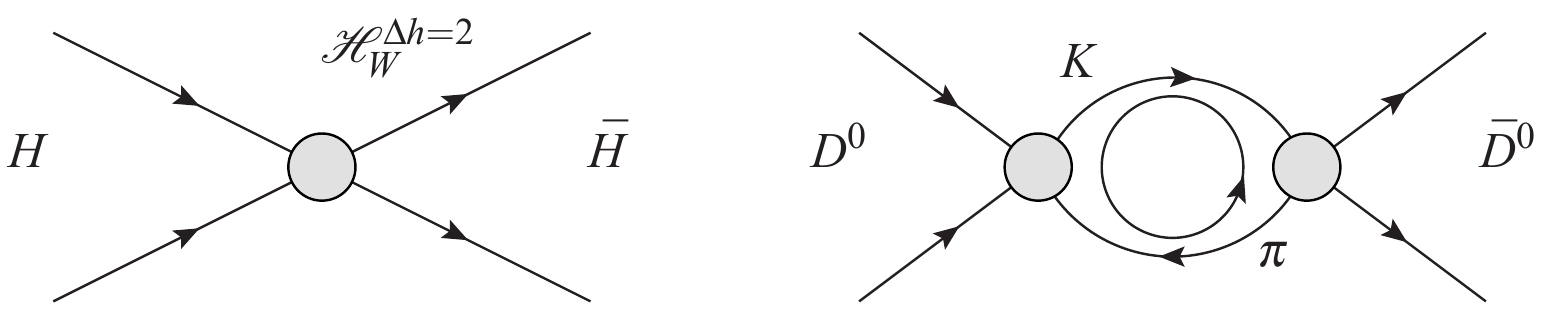}
	\caption{Left: Short-distance diagram. \quad Right: Long-Distance Diagram for $D$-meson mixing.}
	\label{mixingdiagram}
\end{figure}

The hadronic bound states are goverened by strong coupling QCD dynamics. The short-distance interactions are due to the Standard Model (SM) electroweak interactions and possibly Beyond Standard Model (BSM) extensions. The separation of scales allows the high-energy effects to be integrated out and collected into Wilson coefficients. The left panel of Fig.~\ref{mixingdiagram} illustrates the short-distance contributions to neutral meson mixing in terms of effective four-quark operators.

The right panel of Fig.~\ref{mixingdiagram} illustrates the long-distance contribution to neutral meson mixing; between the two $\mathcal{H}_W^{\Delta h=1}$ are insertions of intermediate light-mesons bound states. The two interaction points shown in Fig.~\ref{mixingdiagram} are separated by length scale $1/\Lambda_{QCD}$. Currently, theoretical predictions of the long-distance contribution vary greatly~\cite{hep-ex/9908021}, while current lattice techniques are not yet suited to tackle this problem.

The $\Delta c=2$ and $\Delta b=2$ four-quark operators that are invariant under Lorentz, Fierz and C, P, T transformations are commonly expressed in the following basis,
\begin{equation}
\mathcal{O}_1=\bar{h}^\alpha \gamma^\mu L q^\alpha \bar{h}^\beta \gamma^\mu Lq^\beta, \quad\quad \mathcal{O}_2=\bar{h}^\alpha L q^\alpha \bar{h}^\beta L q^\beta, \quad\quad \mathcal{O}_3=\bar{h}^\alpha L q^\beta \bar{h}^\beta L q^\alpha,
\end{equation}
\begin{equation}
\mathcal{O}_4=\bar{h}^\alpha L q^\alpha \bar{h}^\beta R q^\beta, \quad\quad \mathcal{O}_5=\bar{h}^\alpha L q^\beta \bar{h}^\beta R q^\alpha,
\end{equation}
where $L$ and $R$ are the left and right projection operators. We denote a heavy quark (charm or bottom) as $h$, and a light or strange valence quark as $q$. Operators $\mathcal{O}_4$ and $\mathcal{O}_5$ appear only in BSM models.

The mass differences in the $B_d$ and $B_s$ systems are measured to sub-percent accuracy, and the $B_s$ width difference is measured with 15\% precision~\cite{1207.1158}. The SM prediction of the mass and width differences are dominated by the short-distance contributions in Eq.~(\ref{mixingH}) while in the SM the mass difference depends on $V_{td}$ and $V_{ts}$. Hence, sufficiently accurate lattice QCD calculations of the hadronic matrix elements of the operators that contribute to the SM can be used to test CKM unitarity. In addition, knowledge of the matrix elements of all five operators provides constraints on BSM theories~\cite{1212.5470}.

In the SM, the short-distance contribution to $D$-meson mixing is expected to be much smaller than the long-distance contribution due to GIM suppression between the strange and down quarks, and $\left|V_{ub}V_{cb}^*\right|^2$ suppression of the bottom quark loop. BSM contributions may enhance the short-distance contributions. The current HFAG $D$-mixing experimental averages for the mixing parameters $x=M_{12}/{\Gamma}$ and $y=\Gamma_{12}/2\Gamma$ are measured with respectively 28\% and 17\% precision~\cite{1207.1158}, where $\Gamma$ is the average decay width of the $D^0$ mass eigenstates. Results from next generation flavor factories promise to provide more accurate experimental measurements, which in conjunction with knowledge of the hadronic matrix elements of the five short-distance operators allows for model discrimination between BSM theories~\cite{0705.3650}.

\section{Correlator Analysis}
We use the MILC collaboration's $N_f=2+1$ asqtad ensembles at four different lattice spacings with four or five different light valence sea quark masses per lattice spacing (except for the finest lattice spacing). For each sea-quark-mass ensemble we generate asqtad light quarks with seven or eight different masses. The charm and bottom quarks are simulated with the Fermilab interpretation of the Sheikholeslami-Wohlert action~\cite{0903.3598}.

The lattice formulation for $D$- and $B$-meson mixing are the same, since we use the same action for the valence charm and bottom quarks. The two- and three-point correlation functions we need to extract the hadronic matrix elements are,
\begin{equation}
C^{2pt}(t)=\sum_{\vec{x}}\left<\bar{\chi}(\vec{x},t)\chi(0)\right>, \quad\quad C^{3pt}_i(t_1,t_2)=\sum_{\vec{x}_1,\vec{x}_2}\left<\chi(\vec{x}_2,t_2)\mathcal{O}_i(0)\chi(\vec{x}_1,t_1)\right>,
\end{equation}
where $\chi\equiv\bar{q}\gamma_5 h$ is the meson creation operator.

\begin{table}[h]
\centering
\tabcolsep=0.11cm
\footnotesize
\begin{tabular}{c|c|c|c|c|c}
\toprule
$a (fm)$ & $\left(\frac{L}{a}\right)^3\times \frac{T}{a}$ & $m_l/m_s$ & $am_q$ & $\kappa_b$ & $\kappa_c$\\
\midrule
0.12  &  $24^3\times 64$ & $0.1$ & $0.0050,0.0070,0.0100,0.0200,0.0300,0.0349,0.0415,0.0500$ & $0.0901$ &$0.1254$ \\
0.12  &  $20^3\times 64$ & $0.14$ & $0.0050,0.0070,0.0100,0.0200,0.0300,0.0349,0.0415,0.0500$ & $0.0901$ & $0.1254$\\
0.12  &  $20^3\times 64$ & $0.2$ & $0.0050,0.0070,0.0100,0.0200,0.0300,0.0349,0.0415,0.0500$ & $0.0901$ &$0.1254$\\
0.12  &  $20^3\times 64$ & $0.4$ & $0.0050,0.0070,0.0100,0.0200,0.0300,0.0349,0.0415,0.0500$ & $0.0918$ &$0.1259$\\
\midrule
0.09 & $64^3\times 96$ & $0.05$ & $0.00155,0.0031,0.0062,0.0093,0.0124,0.0261,0.0310$ & 0.0975&0.1275\\
0.09 & $40^3\times 96$ & $0.1$ & $0.0031,0.0047,0.0062,0.0093,0.0124,0.0261,0.0310$ & 0.0976& 0.1275\\
0.09 & $32^3\times 96$ & $0.14$ & $0.0031,0.0047,0.0062,0.0093,0.0124,0.0261,0.0310$ & 0.0977& 0.1275\\
0.09 & $28^3\times 96$ & $0.2$ & $0.0031,0.0047,0.0062,0.0093,0.0124,0.0261,0.0310$ & 0.0979& 0.1276\\
0.09 & $28^3\times 96$ & $0.4$ & $0.0031,0.0047,0.0062,0.0093,0.0124,0.0261,0.0310$ & 0.0982& 0.1277\\
\midrule
0.06 & $64^3\times 144$ & $0.1$ & $0.0018,0.0025,0.0036,0.0054,0.0072,0.0160,0.0188$ & 0.1052& 0.1296\\
0.06 & $56^3\times 144$ & $0.14$ & $0.0018,0.0025,0.0036,0.0054,0.0072,0.0160,0.0188$ & 0.1052& 0.1296\\
0.06 & $48^3\times 144$ & $0.2$ & $0.0018,0.0025,0.0036,0.0054,0.0072,0.0160,0.0188$ & 0.1052& 0.1296\\
0.06 & $48^3\times 144$ & $0.4$ & $0.0018,0.0025,0.0036,0.0054,0.0072,0.0160,0.0188$ & 0.1048& 0.1295\\
\midrule
~0.045 & $64^3\times 192$ & $0.2$ & $0.0018,0.0028,0.0040,0.0056,0.0084,0.0130,0.0160$ & 0.1143& 0.1310\\
\bottomrule
\end{tabular}
	\caption{The simulation parameters and ensemble properties.}
	\label{MILCconfig}
\end{table}

We perform simultaneous Bayesian fits to the two- and three-point correlator data to extract the mixing matrix elements. The fit functions have the same form for both the $D$- and $B$-meson mixing analyses,
\begin{equation}
C^{2pt}\left(t\right)=\sum_{n=0}^{N_s}\left(-1\right)^{n\left(t+1\right)}\frac{\left| Z_n\right|^2}{2E_n}\left(e^{-E_nt}+e^{-E_n\left(T-t\right)}\right),
\label{2pt}
\end{equation}
\begin{equation}
C^{3pt}_i\left(t_2,t_1\right)=\sum_{m,n=0}^{N_s}\left(-1\right)^{n\left(t_2+1\right)}\left(-1\right)^{m\left(\left|t_1\right|+1\right)}\frac{\left<n\left|\mathcal{O}_i\right|m\right>Z_n^\dagger Z_m}{4E_nE_m}e^{-E_nt_2}e^{-E_m\left|t_1\right|}.
\label{3pt}
\end{equation}
In Eqs.~(\ref{2pt}) and (\ref{3pt}), the states with odd $n$, which oscillate in $t$, are of opposite parity.

\begin{figure}[b]
	\centering
  \resizebox{\textwidth}{!}{\begin{minipage}{\textwidth}
		\includegraphics[width=\textwidth]{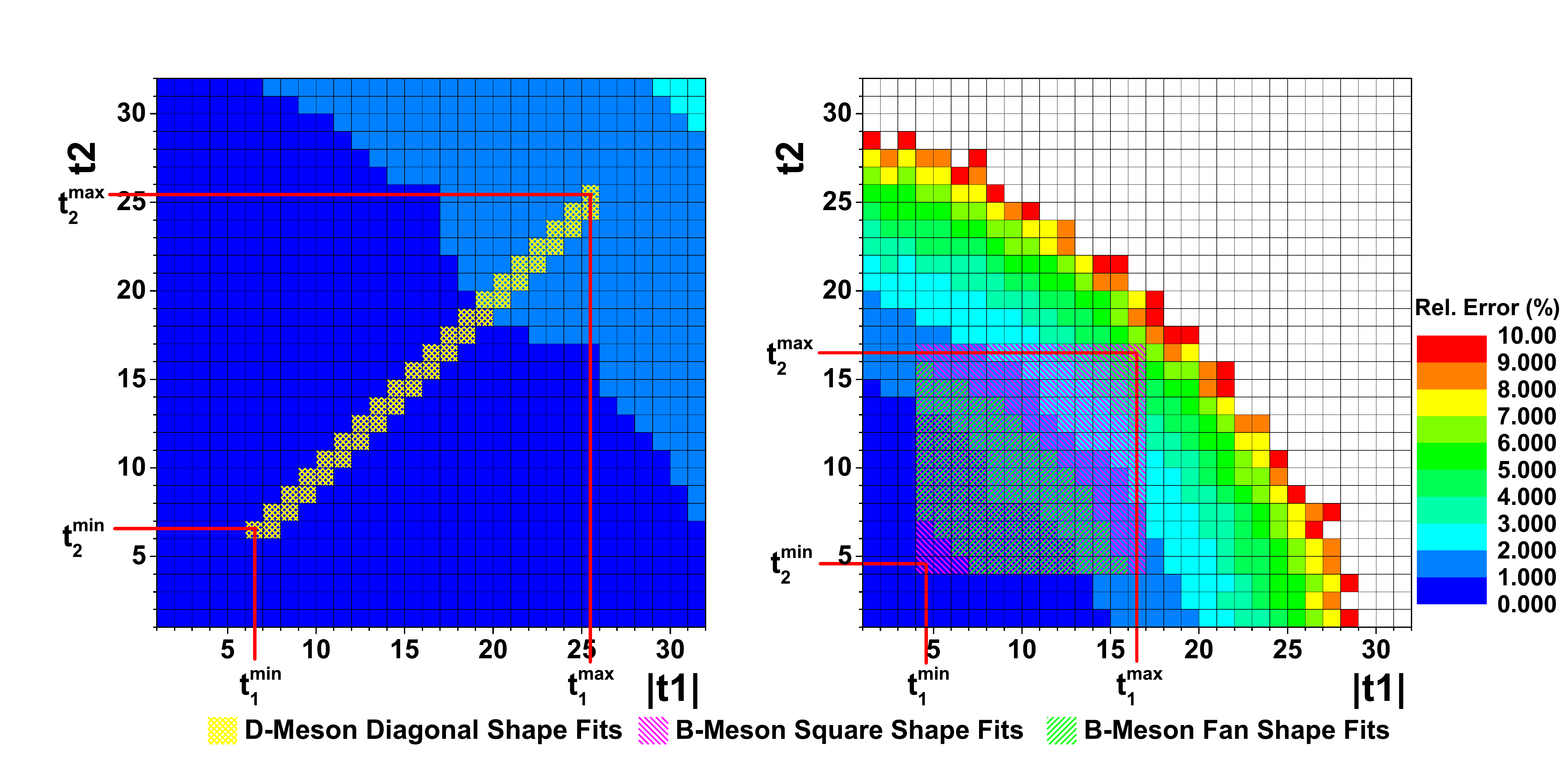}
	\caption{$D$-meson (left) and $B$-meson (right) three-point function relative errors in the ($t_1$, $t_2$)-plane overlaid with the shapes for the fit regions.  
The relative errors are shown as blue-to-red color variation. The fit regions are indicated by the overlaid shaded areas. $D$-meson (left): yellow diagonal. $B$-meson (right): pink (green) for the square (fan) shape. }
	\label{FitRegions}
	\end{minipage}
	}
\end{figure}

Following the method outlined in Ref.~\cite{1112.5642}, we select $N_s=3$ for all $D$- and $B$-meson mixing fits. Thus, we include two states of normal parity and two of opposite parity. The fit region across all ensembles are chosen to have consistent $t_{min}$ and $t_{max}$ values in physical units, keeping correlations between partially quenched data intact. The three-point data included in the fits are restricted to $t_{max} < T/2$, where $T$ is the temporal extent of the lattice, in order to suppress periodic boundary condition effects. Fig.~\ref{FitRegions} shows that for the same ensemble, $B$-meson mixing data exhibits a relatively poor signal-to-noise ratio; fitting over a square or fan shaped regions allows us to extract as much information as possible given this limitation. The $D$-meson mixing signal is much less noisy because the charm quark is less heavy. This permits fits over larger time separations, where the ground state signal dominates. In fact, the number of configurations in each ensemble limits the number of eigenvalues in the covariance matrix that can be resolved. After testing various options~\cite{0803.1909}, we find that fits over two diagonal slices are optimal.

\begin{figure}[b]
	  \centering
		\includegraphics[width=0.9\linewidth]{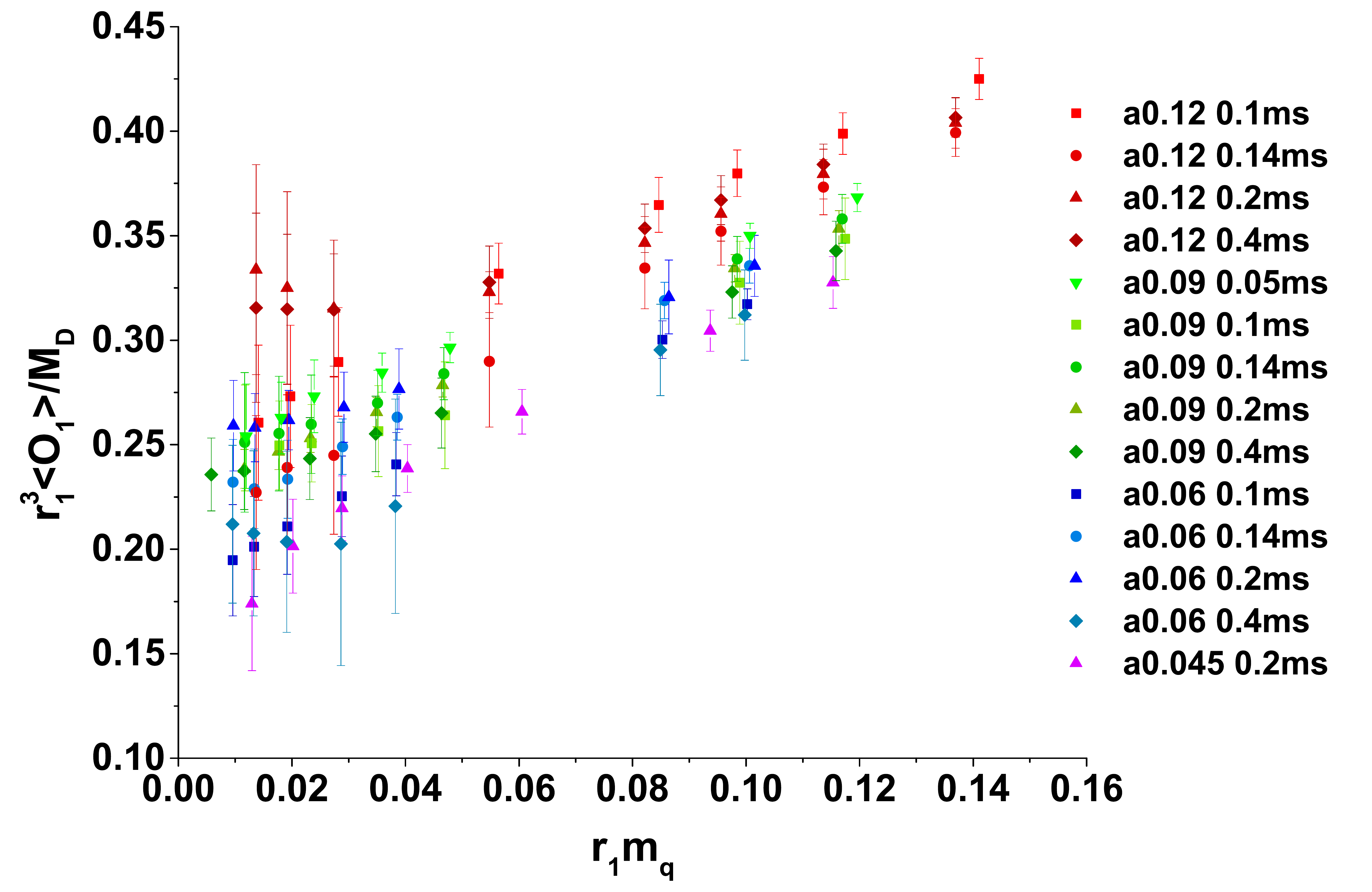}
	  \caption{$D$-mixing: Renormalized $\left<\mathcal{O}_1\right>$ data as a function of light valence quark mass ($m_q$) in $r_1$ units. Statistical errors only. Color is used to indicate the lattice spacing, while shading and shapes indicate the light-to-strange quark mass ratio, as detailed in the legend.}
	  \label{fig:DMixResults}
\end{figure}

With the three-point data in hand, we use one-loop matching to the continuum $\overline{\rm MS}-$NDR scheme.  The matrix elements mix under renormalization,
\begin{equation}
\langle H | \mathcal{O}_i|\bar{H}\rangle^{\rm \overline{MS}-NDR}(m_h) =  \sum_{j=1}^5 \big[ \delta_{ij}+\alpha_s(q^*) \zeta_{ij}(am_h)\big] \langle H | \mathcal{O}_j|\bar{H}\rangle^{\rm lat} ,
\label{eq-renorm}
\end{equation}
where the matching coefficients $\zeta_{ij}$ give the difference between the one-loop lattice and continuum renormalizations, $m_h$ sets the scale, and $\alpha_s(q^*)$ is determined as in Ref.~\cite{0911.5432}. We use the short-hand $\langle {\mathcal O}_i \rangle \equiv \langle H | {\mathcal O}_i|\bar{H}\rangle^{\rm \overline{MS}-NDR}(m_h)$ to denote the renormalized matrix elements. Using Eq.~(\ref{eq-renorm}), we obtain the renormalized $D$- and $B$-meson mixing matrix elements of all five operators on all ensembles. Our results for $\left<{\mathcal O}_1\right>$ are shown as an example in Fig.~\ref{fig:DMixResults} and \ref{fig:BMixingO1} as a function of  the light valence-quark mass.

\section{Chiral and Continuum Extrapolation}
The chiral-continuum extrapolation of the mixing matrix elements is guided by SU(3) partially-quenched, heavy-meson, staggered chiral perturbation theory (PQHMS$\chi$PT)~\cite{1303.0435}, which includes the dominant light-quark discretization effects from taste violations. These violations have the usual effect on the chiral logs, and for neutral heavy-meson mixing they also generate terms that mix the $\langle \mathcal{O}_i \rangle$ with each other~\cite{1303.0435}. The five matrix elements $\langle \mathcal{O}_{1\dots 5} \rangle$ form a complete basis, introducing no new low-energy constants.  Mixing occurs among  $\langle \mathcal{O}_1 \rangle$, $\langle \mathcal{O}_2 \rangle$, $\langle \mathcal{O}_3 \rangle$, and, separately, among $\langle \mathcal{O}_4 \rangle$, $\langle \mathcal{O}_5 \rangle$. For example, the expression for $\left<B_q^0\left|\mathcal{O}_1\right|\bar{B}_q^0\right>$ at next-to-leading order in PQHMS$\chi$PT is
\begin{eqnarray}
	\langle \overline{B}_q^0|{O}_1^q|B_q^0 \rangle =&&\beta_1\Bigg( 1+\frac{{W}_{q\overline{b}}+{W}_{b\overline{q}}}{2} + T_q  + Q_q +\tilde{T}^{(\rm{a})}_q+\tilde{Q}^{({\rm a})}_q\Bigg)\nonumber \\
	&& + (2\beta_2+2\beta_3) \tilde{T}^{(\rm{b})}_q + (2\beta'_2+2\beta'_3) \tilde{Q}^{({\rm b})}_q + \mbox{NLO analytic terms}.
	\label{eq:chipt}
\end{eqnarray}
The terms $W$, $T$, and $Q$, are ``correct spin'' contributions from wave-function renormalization, tadpole diagrams, and sunset diagrams, respectively. The terms $\tilde{T}$ and $\tilde{Q}$, again from tadpole and sunset diagrams, are the contributions from wrong-spin operators. The $\beta_i$ and $\beta'_i$ are the leading-order LEC's for the matrix element $\left<B_q^0\left|\mathcal{O}_i\right|\bar{B}_q^0\right>$. We are currently implementing simultaneous fits to $\left\{\left<\mathcal{O}_1\right>,\left<\mathcal{O}_2\right>,\left<\mathcal{O}_3\right> \right\}$ and $\left\{\left<\mathcal{O}_4\right>,\left<\mathcal{O}_5\right> \right\}$ to handle the chiral extrapolation cleanly.

\section{Status and Conclusions}
The correlator analysis for both the $D$- and $B$-meson mixing projects is final. However, in the $D$-meson case, the renormalization and matching of the BSM operators is still in progress. We are currently finalizing the $B$-meson chiral-continuum extrapolation for all five operators.

\begin{figure}[b]
	\centering
		\includegraphics[width=0.9\linewidth]{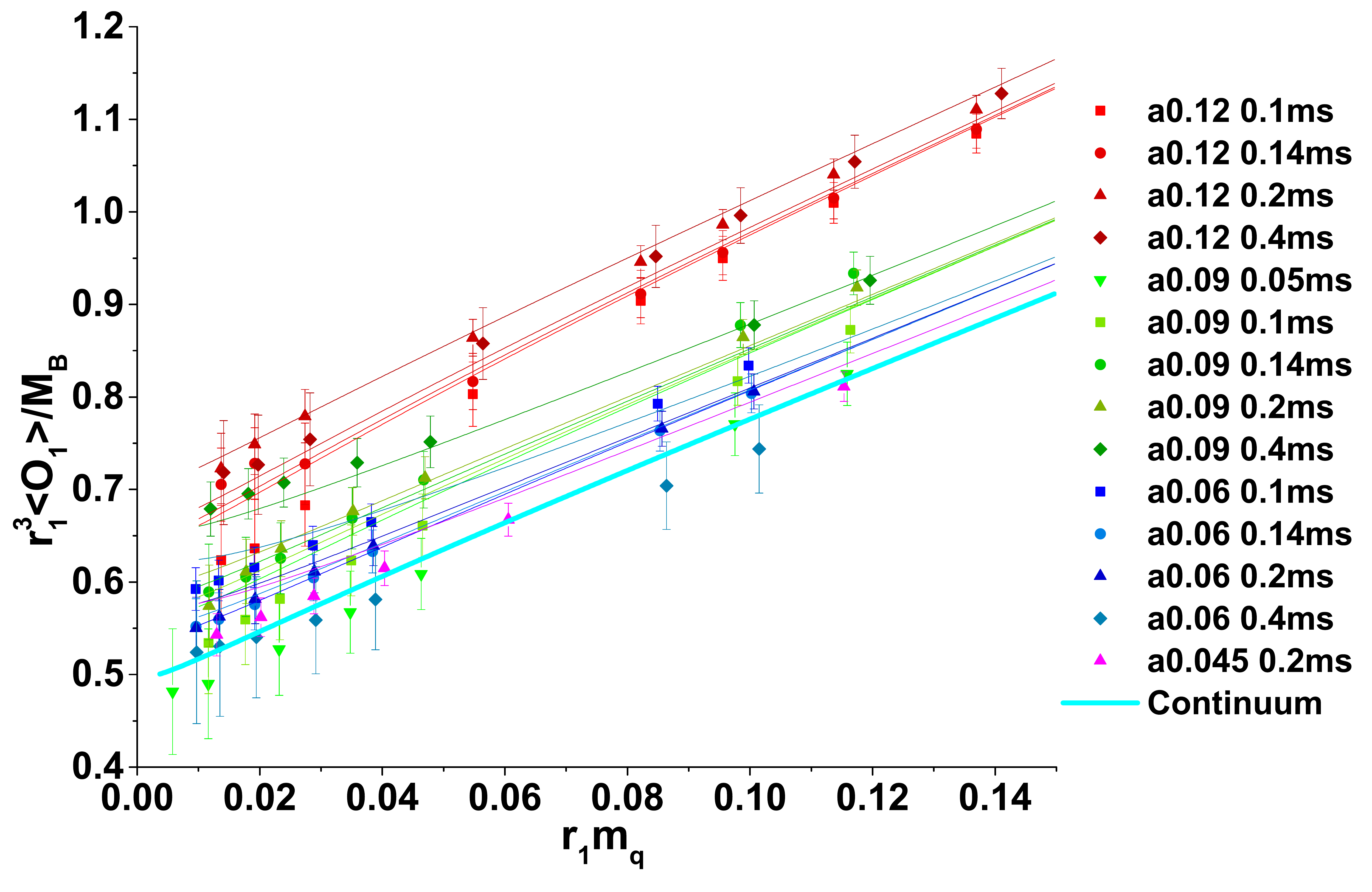}
	\caption{$B$-mixing: Renormalized $\left<\mathcal{O}_1\right>$ data as a function of $r_1 m_q$ together with a preliminary chiral fit and continuum extraploation using Eq.~(\protect\ref{eq:chipt}) plus NNLO analytic terms~\protect\cite{1112.5642}. Statistical errors only. The color scheme is the same as Fig.~\protect\ref{fig:DMixResults}.}
	\label{fig:BMixingO1}
\end{figure}

In the $B$-meson mixing case, we use the fan shape for the correlators  of $\mathcal{O}_1$, $\mathcal{O}_2$, and $\mathcal{O}_3$, and the square shape for the correlators of $\mathcal{O}_3$, $\mathcal{O}_4$, and $\mathcal{O}_5$.  Hence two sets of  results for $\left<\mathcal{O}_3\right>$ serve as a cross-check of our fit methods. Preliminary $B$-mixing chiral fits are shown for $\langle {\mathcal O}_1 \rangle$ in Fig.~\ref{fig:BMixingO1}. While not included in the proceedings, preliminary chiral-continuum fits also exist for the others. Effects from heavy-light meson flavor- and hyperfine-splittings~\cite{1112.3051} and finite volume (expected to be $< 1\%$) are not yet included in Fig.~\ref{fig:BMixingO1}.

We expect that our final error budget for the $B$-meson mixing analysis will be consistent with the expectations presented in Ref.~\cite{1112.5642}. In particular, we do not expect  the complications due to the mixing with "wrong-spin" terms to have a significant effect on the chiral-continuum extrapolation error. Indeed, we expect a reduction of the dominant systematic errors in our final analysis due to the inclusion of ensembles with finer lattice spacings and smaller light sea-quark masses. Once our analyses are final, we will present our results for all five matrix elements and for combinations of matrix elements that are useful for phenomenology.

\section*{Acknowledgements}
This work was supported by the U.S. Department of Energy, the National Science Foundation, the Universities Research Association,  the MINECO, Junta de Andaluc\'ia, and the European Commission. Computation for this work was done at the Argonne Leadership Computing Facility (ALCF), the National Center for Atmospheric Research (UCAR), the National Center for Supercomputing Re- sources (NCSA), the National Energy Resources Supercomputing Center (NERSC), the National Institute for Computational Sciences (NICS), the Texas Advanced Computing Center (TACC), and the USQCD facilities at Fermilab, under grants from the NSF and DOE.

{

}
%\end{multicols}

\end{document}